# How Far Can We Trust Chaos?
# Extending the Horizon of Predictability

Alexandros K. Angelidis [1, a], Georgios C. Makris [2, b], Evangelos Ioannidis [3, c], Ioannis E. Antoniou [2, d], Charalampos Bratsas [1, e]

[1] Department of Information and Electronic Engineering, International Hellenic University, Thessaloniki, 574 00, Greece; a.angelidis@ihu.gr; cbratsas@ihu.gr

[2] Department of Mathematics, Aristotle University of Thessaloniki, Thessaloniki, 541 24, Greece; geormakr@csd.auth.gr; iantonio@math.auth.gr

[3] Department of Economics, Aristotle University of Thessaloniki, Thessaloniki, 541 24, Greece; ioannide@econ.auth.gr

[a] ORCID: 0009-0003-0777-6880
[b] ORCID: 0000-0003-0246-4258
[c] ORCID: 0000-0001-6923-4700
[d] ORCID: 0000-0001-8080-0945
[e] ORCID: 0000-0001-6400-3233

**Abstract**

Chaos reveals a fundamental paradox in the scientific understanding of Complex Systems. Although chaotic models may be mathematically deterministic, they are practically non-determinable due to the finite precision, which is inherent in all computational machines. Beyond the horizon of predictability, numerical computations accumulate errors, often undetectable. We investigate the possibility of reliable (error-free) time series of chaos. We prove that this is feasible for two well-studied isomorphic chaotic maps, namely the Tent map and the Logistic map. The generated chaotic time series have unlimited horizon of predictability. A new linear formula for the horizon of predictability of the Analytic Computation of the Logistic map, for any given precision and acceptable error, is obtained. Reliable (error-free) time series of chaos serve as "gold standard" for chaos applications. The practical significance of our findings include: **(i)** the ability to compare the performance of neural networks that predict chaotic time series, **(ii)** the reliability and numerical accuracy of chaotic orbit computations in encryption, maintaining high cryptographic strength, and **(iii)** the reliable forecasting of future prices in chaotic economic and financial models.

## 1. Introduction

Chaos, discovered by Poincaré at the end of the 19th century, emerged as analytic non-integrability and dynamical instability of the three-body problem [1]. Later, Lorenz faced the computational implications of chaos and used the term "Butterfly Effect" to describe how tiny changes in initial conditions can lead to vastly divergent outcomes [2], [3], [4]. Sensitive dependence on initial conditions [5], [6], [7], [8], [9], [10], [11], [12] has been observed in a wide range of fields, from physics and engineering to biology, economics, sociology, geology, and astronomy. Chaos is known to impose intrinsic limitations to predictability. The horizon of predictability was introduced as the duration $\tau_P$ of reliable predictions within a prescribed accuracy by Lighthill in 1986 [13] who *"wished to apologize for having misled the general educated public by spreading ideas about the determinism of systems satisfying Newton's laws of motion that, after 1960, were to be proved incorrect"*. Liao used later (2009) the name "critical predictable time", as the last time-step that computed chaotic solutions remain reliable [14], which is precisely the horizon of predictability. A computation is considered reliable, if the precision of successive iterations remains within acceptable bounds.

The horizon of predictability $\tau_P$ depends on the specific dynamical model, the initial condition (state or seed), the simulation algorithm, the acceptable error $\varepsilon$ and the computation accuracy $\delta$. Almost all mainstream programming languages and software use floating-point arithmetic (IEEE 754) with precision (mantissa) $m = 32$-bit or $m = 64$-bit (single precision or double precision respectively). The accuracy $\delta$ is related with the mantissa of $m$ decimal digits or with $\log_2(10)\, m \simeq 3.32m$ binary digits (bits):

$$\delta = 10^{-m} = 2^{-log_2(10)m} \simeq 2^{-3.32m} \tag{1}$$

The acceptable error $\varepsilon$ cannot be greater than the computation accuracy $\delta$: $\varepsilon \leq \delta$. In decimal representation this translates to $k \leq m$, with $k$ denoting the acceptable precision:

$$\varepsilon = 10^{-k} \tag{2}$$

It is conventionally accepted that meaningful predictions cannot be made over a duration of more than two or three times the Lyapunov time $\tau_L$, which is the inverse of the system's Maximal Lyapunov exponent $\lambda$: $\tau_L = \frac{1}{\lambda}$. Assuming local exponential divergence of nearby orbits, the Lyapunov time is the duration for distance growth by a factor $e \simeq 2.72$. The Lyapunov time is crucial in forecasting tasks, because it mirrors the limits of predictability of a chaotic system and allows to fairly compare the predictive accuracy in different dynamical systems [15].

Assuming local exponential divergence of nearby orbits, the horizon of predictability $\tau_P$ is related to the Lyapunov time $\tau_L$ by the formula:

$$\tau_P \simeq \tau_L \ln\left(\frac{\varepsilon}{\delta}\right) \tag{3}$$

Eq. (3) follows from the fact that at the horizon of predictability $\tau_P$, the initial accuracy $\delta$ has grown to the value of the acceptable error $\varepsilon$: $\varepsilon \simeq \delta e^{\lambda \tau_P}$.

The horizon of predictability $\tau_P$ can also be expressed in terms of the Kolmogorov-Sinai entropy production rate $\hbar$ [16], [17], [18]. According to Pesin's formula [19], [20] the entropy production is the sum of positive Lyapunov exponents. For one dimensional chaotic maps [21], [22]:

$$\hbar = \frac{1}{\ln 2}\lambda \tag{4}$$

Therefore, the horizon of predictability is inversely proportional to the entropy production rate:

$$\tau_P \simeq \frac{1}{\hbar \cdot \ln 2} \ln\left(\frac{\varepsilon}{\delta}\right) \tag{5}$$

Chaos has captivated scientists with its intrinsic unpredictability. Yet behind this fascination lies a paradox that continues to haunt computational science: although chaos is deterministic, it is not determinable. As digital computers remain tethered to finite precision arithmetic, simulating chaos accurately over extended iterations becomes not just problematic, it becomes, in principle, impossible. The limitations and dangers of simulating chaotic systems using digital computers have been discussed extensively [23], [24], [25], [26], [27], [28], [29], [30], [31], [32], [33]. Converting mathematical models to executable code for computation does not avoid the inevitable computational error. Chaotic systems quickly amplify small numerical inaccuracies into tremendously large ones in the long run. Thus, computational scenaria are actually hallucinations generated by the limited mantissa of the computer, representing a finite subset of rational numbers $\tilde{\mathbb{Q}}$ [34]. As the required memory increases exponentially, when the number of digits exceeds the mantissa, the numerically computed chaotic behavior does not represent the behavior of the system, but incorporates the numerical errors imposed by the computation process beyond the horizon of predictability [35], [36], [37], [38], [39], [40], [41], [42], [43], [44], [45]. Of course, the simulations are reliable and therefore meaningful for a number of steps not greater than the horizon of predictability. Therefore, the estimation of the horizon of predictability is unavoidable for the estimation of reliable computations of chaotic systems. Beyond the horizon of predictability, evolution is effectively described as a stochastic process using probabilistic properties of chaos [46], [47], [48], [49], [50], [51], [52], [53], [54], [55], [56], [57], [58], [59], [60], [61], [62], [63], [64], [65]. The horizon of predictability specifies actually the bound of rationality, a concept introduced by Herbert Simon in the context of economic and social sciences [66], [67].

The mathematical theory of dynamical systems is incomplete with respect to the question of computability. Not only integrability is undecidable, but also actual integration is not an algorithmically computable problem [68], [69]. Therefore, there is no systematic method for reliable computation of chaos. However, one cannot a priori exclude the possibility for reliable computation of specific chaotic systems.

The goal of this work is to explore the possibility for reliable computation of some specific chaotic systems in order to find the horizons of predictability associated with different simulations. We demonstrate that this goal is feasible for two simple isomorphic chaotic maps, namely the Tent map and the Logistic map. Such reliable computations of chaos actually serve as "gold standard" for chaos applications. In specific, the practical significance of our findings include: (i) the ability to compare the performance of neural networks that predict chaotic time series, (ii) the reliability and numerical accuracy of chaotic orbit computations in encryption, and (iii) the reliable forecasting of future prices in chaotic economic and financial models.

The structure of this paper is the following. After a review of Computations of Chaos (Section 2) and in particular of the Logistic map (Section 3**)**, we discuss the usual Recursive Computation (Section 4) and the Analytic Computation (Section 5) using the formula found by Ulam and von Neumann [70], [71], [72], [73], [74]. Moreover, we introduce and validate a new linear formula to estimate the horizon of predictability of the Analytic Computation of the Logistic map, for any given precision and acceptable error (Proposition 1). The horizons of predictability of both computations are effectively indistinguishable and are compared using the Error-Free Computation of the Logistic map (Section 6). This is possible using the isomorphism (conjugation) of the Logistic map with the Tent map which is also computable using rational representation (Appendix A). The concluding remarks are presented in Section 7.

## 2. Computation of Chaos

The difficulties of simulating chaos computations have been discussed by several authors.
Lorenz in 1989 [3] found that chaotic behavior in numerical simulations can arise not only from the underlying dynamics, but also from the numerical method used, a phenomenon he called "computational chaos." He demonstrated that such artifacts could cause bifurcations and the onset of chaos in discrete approximations, even when the system does not exhibit such features. Moreover, Lorenz described in 2006 [4] "computational periodicity", the opposite phenomenon, where numerical solutions appear periodic even though the orbit is chaotic.

Corless et al. in 1991 [75] showed that beyond causing numerical instabilities, when using large step sizes some numerical methods can suppress genuine chaos in continuous dynamical systems and introduce false stability. They also noted the issue of spurious apparent chaos arising from roundoff errors. They argued that this problem must be kept in mind, and efforts to ensure that it does not happen in the numerical solution of the differential equations must be made. Moreover, Corless in 1994 [76] critically examined the reliability and limitations of numerical simulations in capturing the behavior of chaotic systems. He argued that while simulations are often used to analyze chaos, such as through computed orbits or Lyapunov exponents, numerical methods can either introduce artificial chaos or suppress real chaotic behavior due to discretization, round-off errors, and algorithmic choices.

Yao in 2010 [77] raised concerns about the reliability of numerical solutions for chaotic nonlinear differential equations. He argued that the use of discrete numerical methods and finite computer precision introduces errors that contaminate the results and said that this is not just a technical issue. Chaotic systems have properties, like unstable manifolds and virtual separatrices, which amplify small numerical errors. These errors grow along unstable directions, violating Von Neumann stability criterion [78], needed to ensure that numerical methods converge to the correct solution. Yao emphasized that although numerical solutions may appear reasonable, they may be fundamentally flawed and should not be trusted as valid representations of the true behavior of the system. He also concluded that numerical solutions to certain chaotic equations (Lorenz, Rössler, and Kuramoto-Sivashinsky) are inherently unreliable due to fundamental limitations in the way we compute them.

Lozi in 2013 [79] emphasized that although numerical methods have significantly advanced our ability to visualize and analyze chaotic systems, they come with serious limitations that require caution. The core issue lies in the intrinsic sensitivity of chaotic systems to initial conditions and the finite precision of numerical computations. This can lead to misleading results, such as pseudo-chaotic behaviors or long transient orbits that eventually collapse into stable periodic cycles. Even popular systems like the Lorenz and Hénon models can produce results that misrepresent true dynamical behavior when simulated.

Boghosian et al. in 2019 [29] demonstrated a systematic numerical error in the simulation of chaotic nonlinear dynamical systems and specifically, of the generalized Bernoulli map. This error differs from the typical rounding or precision errors and would remain for any finite-precision mantissa, however large. It arises from the discreteness of floating-point numbers, their non-uniform distribution along the real axis, and their inability to represent points on periodic orbits of the dynamics in a precise way, leading to a severely truncated spectrum of these orbits and distorted statistical properties. While the paper examines a relatively simple model, the authors caution that more complex systems, like turbulent fluid flows or molecular dynamics, are likely to suffer from even greater inaccuracies.

Qin and Liao in 2020 [80] argued that numerical simulations in a long interval of time should be checked very carefully because the numerical noises (i.e., truncation error and round-off error) can lead to both quantitatively and qualitatively, huge deviation of the spatio-temporal chaotic system not only in orbits but also in statistics.

Coveney in 2024 [33] explored the limitations of simulating chaotic systems on digital computers due to the discrete nature of floating-point number. Coveney argues that this limitation is rooted in Sharkovskii's theorem [81], [82], which mathematically establishes that finite number systems cannot capture the full dynamics of chaotic systems.

## 3. Computation of the Logistic Map

The Logistic map is a well-known simple chaotic discrete dynamical system defined by the quadratic iteration map on the unit interval $[0,1]$ [83], [84], [85], [53], [63]:

$$x_{n+1} = S(x_n) = 4x_n(1 - x_n), \qquad n = 0, 1, 2, \ldots \tag{6}$$

The Logistic map is an extremely rare example of exactly solvable chaotic system. The explicit analytic solution was found by Ulam and von Neumann [70], [71], [72], [73], [74]:

$$x_n = \sin^2\left(2^n \arcsin\left(\sqrt{x_0}\right)\right), \qquad n = 1, 2, \ldots \tag{7}$$

Using Eq. (7), we can avoid iterative computation as we have the value of $x$ at each iteration stage directly. Therefore, we may hope to achieve error-free computation using Eq. (7).

The Logistic map is isomorphic (topologically conjugate) with the piecewise linear Tent map on the unit interval $[0,1]$:

$$y_{n+1} = T(y_n) = \begin{cases} 2y_n \ , & y_n < \dfrac{1}{2} \\ 2(1 - y_n) \ , & \dfrac{1}{2} \leq y_n \end{cases} \tag{8}$$

The conjugacy transformation [86], [87], [88] is:

$$x_n = g(y_n) = \sin^2\left(\frac{\pi y_n}{2}\right), \qquad y_n = g^{-1}(x_n) = \frac{2}{\pi}\arcsin\left(\sqrt{x_n}\right) \tag{9}$$

## 4. Recursive Computation of the Logistic Map

The Logistic and the Tent maps are isomorphic [16], [17], [18] as they have the same Lyapunov exponent $\lambda = \ln 2$ [5], [10] and the same Entropy production $\hbar = 1$ (one bit per iteration). Therefore, the error grows exponentially. Saito and Ito (2014) [89] stated that "*even with the most powerful computers available only the first few tens of time steps of the true orbit can be generated for the Logistic map with generic rational parameters and initial values*". Other works detected false periodicity. For instance, Pershon and Povinelli (2012) [90] found that using finite precision floating-point numbers results in periodic orbits. This periodicity was also explored by Galias, 2021 [91] and Klöwer et al., 2023 [92]. Other studies investigated the last significant digits of the Logistic map. Liu et al. (2014) [93] found that the last significant bits which are usually lost due to rounding are important for the randomness and ergodicity of the Logistic map, and, in a more recent study, Valle and Brune (2024) [94] explored the behavior of the least significant digits in the orbits of the Logistic map and found periodicity. Oteo and Ros (2007) [95] conducted a detailed statistical analysis of double precision errors for the Logistic map, observing significant divergence after relatively few iterations. They found that the horizon of predictability of the Recursive Computation of the Logistic map depends linearly on the precision $m$:

$$\tau_P^R \simeq 3.3m \tag{10}$$

The linear dependence of the horizon of predictability on the precision $m$ was also found by Machicao and Bruno (2017) [96], by Wang and Pan (2018) [97], by Nepomuceno and Mendes (2017) [98] and by Peixoto et al. (2018) [99].

## 5. Analytic Computation of the Logistic Map

The Analytic Computation of the Logistic map, Eq. (7), is free from iterative errors. However, in practice, Oteo and Ros (2007) [95] observed that significant round-off errors are accumulated, making Eq. (7) unreliable for practical implementation. Also, Akritas, Antoniou and Ivanov (2000) [100] approached discrete chaotic maps using Chebyshev neural networks, employing the Logistic map's analytic solution to compute orbits over thousands of iterations. They highlighted the practical difficulties and high computational cost associated with such simulations, emphasizing the uncertainty surrounding the reliability of numerically generated chaotic orbits. In Proposition 1 we mathematically confirm these empirical observations regarding Eq. (7).

**Proposition 1.** The horizon of predictability $\tau_P^A$ of the Analytic Computation of the Logistic map Eq. (7), with mantissa *$m$, Eq. (1) and acceptable precision $k$, Eq. (2) is:*

$$\tau_P^A = (m - k) \cdot \log_2(10) \tag{11}$$

**Proof**

Let, $\theta_0 = \arcsin\left(\sqrt{x_0}\right)$. Then the Analytic Computation of the Logistic map, Eq. (7), is written as follows:

$$x_n = \sin^2(2^n \theta_0)\,, \qquad n = 1, 2, \dots$$

Assuming $x_0 \in [0,1]$, then $\sqrt{x_0} \in [0,1]$, so $\theta_0 \in \left[0, \frac{\pi}{2}\right]$.

Since $\theta_0$ is an irrational angle, we use an approximation $\tilde{\theta}_0$ to $m$ decimal digits, with absolute error:

$$\left|\theta_0 - \tilde{\theta}_0\right| \leq 10^{-m}$$

From the **Mean Value Theorem** of the differentiable function $\sin^2$, there exists some $\xi$ in the interval $(\zeta_1, \zeta_2)$, such that:

$$(\sin^2)'(\xi) = \frac{\sin^2(\zeta_2) - \sin^2(\zeta_1)}{\zeta_2 - \zeta_1}$$

for any $\zeta_1 < \zeta_2$ within the interval $\left[0, 2^n \cdot \frac{\pi}{2}\right]$.

We have:

$$\sin^2(\zeta_2) - \sin^2(\zeta_1) = (\sin^2)'(\xi)(\zeta_2 - \zeta_1)$$

Taking the absolute value and using the fact that: $|(\sin^2)'(\xi)| = |\sin(2\xi)| \leq 1$, for all $\xi$, we have:

$$|\sin^2(\zeta_2) - \sin^2(\zeta_1)| \leq |\zeta_2 - \zeta_1|$$

Therefore,

$$|x_n - \tilde{x}_n| = \left|\sin^2(2^n \theta_0) - \sin^2\left(2^n \tilde{\theta}_0\right)\right| \leq \left|2^n \theta_0 - 2^n \tilde{\theta}_0\right| = 2^n \left|\theta_0 - \tilde{\theta}_0\right|$$

As the acceptable error is $\varepsilon = 10^{-k}$, Eq. (2), we have:

$$|x_n - \tilde{x}_n| \leq 2^n \left|\theta_0 - \tilde{\theta}_0\right| \leq 10^{-k} \Rightarrow 2^n \leq \frac{10^{-k}}{\left|\theta_0 - \tilde{\theta}_0\right|}$$

From the last inequality we have:

$$n \leq \log_2\left(\frac{10^{-k}}{\left|\theta_0 - \tilde{\theta}_0\right|}\right) = -k \cdot \log_2(10) - \log_2\left(\left|\theta_0 - \tilde{\theta}_0\right|\right)$$

At the horizon of predictability, we have maximal computation accuracy $\delta = 10^{-m}$, Eq. (1):

$$\tau_P^A = -k \cdot \log_2(10) - \log_2(10^{-m}) = -k \cdot \log_2(10) + m \cdot \log_2(10) = (m - k) \cdot \log_2(10) \qquad \square$$

In other words, the horizon of predictability $\tau_P^A$ of the Analytic Computation of the Logistic map is proportional to $\log_2(10) \cdot m$:

$$\tau_P^A \simeq 3.32m \tag{12}$$

**Remark**

Therefore, the horizon of predictability $\tau_P^A$ of the Analytic Computation is more or less the same with the horizon of predictability $\tau_P^R$ of the Recursive Computation Eq (10). In other words, the Analytic Computation does not improve the Recursive Computation, because the term $2^n$ inside the sine function causes also accumulation of errors, resulting also in exponential growth as $n$ increases (Proposition 1). Both computations are unreliable after a number of iterations beyond the horizon of predictability Eq. (10), Eq. (11), and Eq. (12).

## 6. Error-Free Computation of the Logistic map

The isomorphism $g$, Eq. (9), allows us to simulate the Logistic map by iterating the Tent map which can be computed error-free (Appendix A). The error introduced by the transformation $g$, Eq. (9), is bounded according to:

**Proposition 2**. *The error of the transformation $g$ of the Tent map to the Logistic map, Eq. (9), is bounded for each iteration by the number $\frac{1}{2}\cdot 10^{-m}$, $m$ is the mantissa, Eq. (1)*:

$$|g(y) - \tilde{g}(y)| \leq \frac{1}{2}\cdot 10^{-m} \tag{13}$$

**Proof**:
The transformation function $g(y) = \sin^2\left(\frac{\pi y}{2}\right)$, Eq. (9), involves the product of a rational number $(y)$ with the irrational number $(\pi)$. The number $y$ is given accurately without approximation from the Tent map.
The error of the transformation function $g$ is:

$$|g(y) - \tilde{g}(y)| = \left|\sin^2\left(\frac{\pi y}{2}\right) - \sin^2\left(\frac{\tilde{\pi} y}{2}\right)\right|$$

Where $\tilde{\pi}$ denotes the rational approximation $of$ $\pi$ rounding or truncated in $m$ decimal digits.
Using the argument in the proof of Proposition 1, for the transformation function $g(y) = \sin^2\left(\frac{\pi y}{2}\right)$ we have:

$$|g(y) - \tilde{g}(y)| = \left|\sin^2\left(\frac{\pi y}{2}\right) - \sin^2\left(\frac{\tilde{\pi} y}{2}\right)\right| \leq \left|\frac{\pi y}{2} - \frac{\tilde{\pi} y}{2}\right| = \left|\frac{1}{2}(\pi y - \tilde{\pi} y)\right| = \frac{|y|}{2}|\pi - \tilde{\pi}|$$

As $y \leq 1$ we get:

$$|g(y) - \tilde{g}(y)| \leq \frac{1}{2}|\pi - \tilde{\pi}|$$

Since $\pi$ is approximated to $m$ decimal digits, the absolute error satisfies:

$$|\pi - \tilde{\pi}| \leq 10^{-m}$$

Therefore:

$$|g(y) - \tilde{g}(y)| \leq \frac{1}{2}\cdot 10^{-m}$$

□

To test the practical reliability of this method we used two different precisions $m = 10$ and $m = 100$ to compute the same orbit, with $m = 10$ and $m = 100$ significant digits. In Table 1 we present the values of some iterations of the two orbits and the errors. The only differences appear in the last digits of each step (after 10 significant digits) as predicted by Proposition 2. Fig. 1 shows that the absolute error of the computed orbits never exceeds the bound $\frac{1}{2}\cdot 10^{-10}$ of Proposition 2, for $m = 10$.

**Table 1**. The error of some sample iterations of the Error-Free Computation of the Logistic map using two different precisions $m = 10$ and $m = 100$. For the first 10 significant digits, each point is identical to the one with higher precision. The only difference occurs only after exceeding the 10 significant digits (red numbers). The Computation Error (3rd column of the Table) is slightly lower than the bound $\frac{1}{2}\cdot 10^{-10}$ of Proposition 2, for $m = 10$.

| **Iteration** | **$m = 10$ Digits Precision** | **$m = 100$ Digits Precision** | **Computation Error $\lvert g(y) - \tilde{g}(y)\rvert$** |
|---|---|---|---|
| **1** | 0.03713801469**53** | 0.03713801469**8137270**… | $2.835734753\cdot 10^{-12}$ |
| **10** | 0.8951968882**538** | 0.8951968882**66336955**… | $1.248749013\cdot 10^{-11}$ |
| **100** | 0.9552148109**287** | 0.9552148109**43948346**… | $1.521106768\cdot 10^{-11}$ |
| **1000** | 0.4922984262**666** | 0.4922984262**90442852**… | $2.379963405\cdot 10^{-11}$ |
| **10000** | 0.1316926680**174** | 0.1316926680**22347760**... | $4.925794814\cdot 10^{-12}$ |

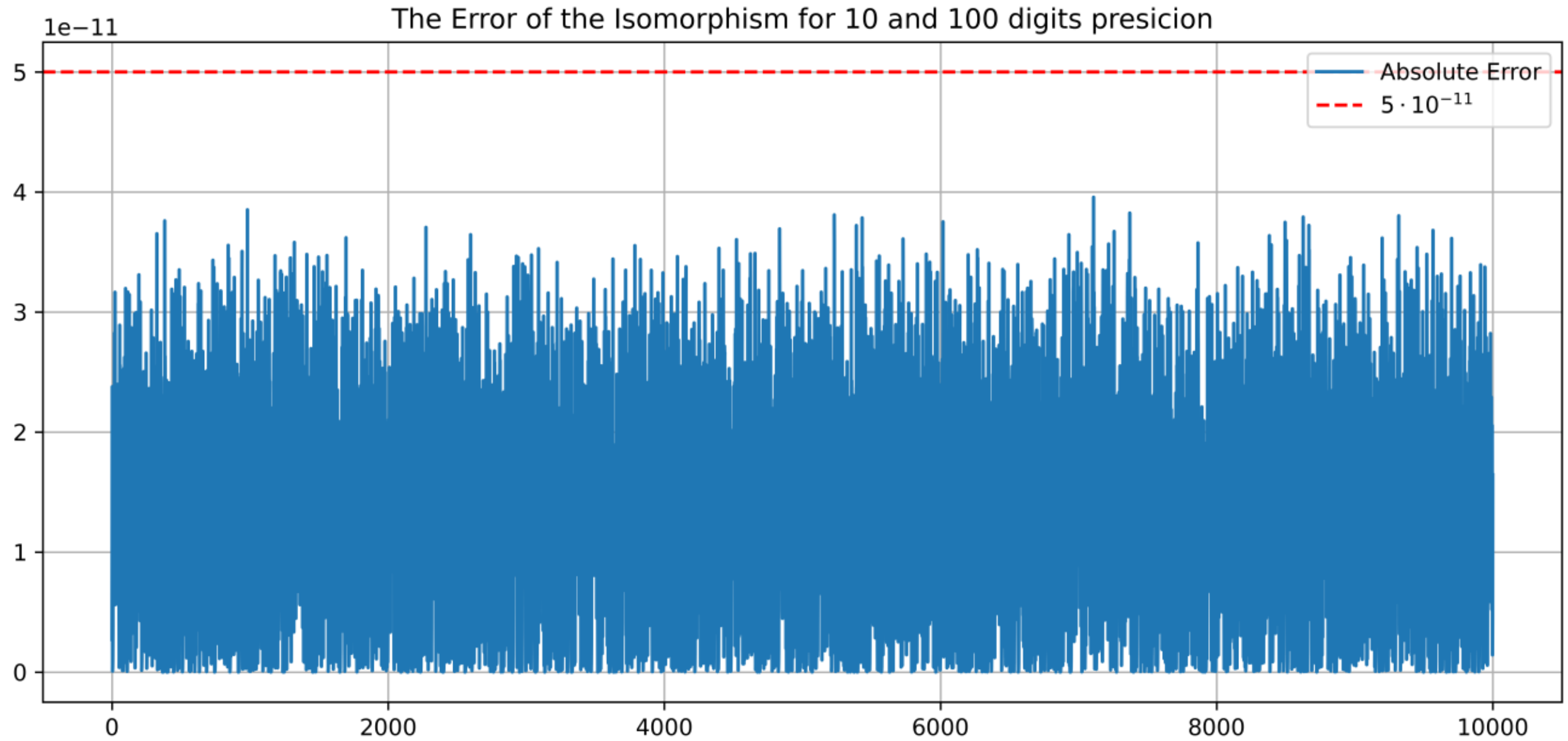

**Fig. 1.** The error of the Error-Free Computation of the Logistic map using two different precisions $m = 10$ and $m = 100$. We observed that for all 10,000 iterations, the error is bounded by the theoretical error $\frac{1}{2} \cdot 10^{-10} = 5 \cdot 10^{-11}$ of Proposition 2. The bound $5 \cdot 10^{-11}$ is depicted as the horizontal red dashed line.

We compare the Analytic Computation and the Recursive Computation of the Logistic map with the Error-Free Computation via the Tent map in Table 2. The acceptable error, Eq. (2), is $\varepsilon = 10^{-2}$. The horizon of predictability is the time step at which the error first exceeds the acceptable error $\varepsilon$. The results for acceptable errors $\varepsilon = 10^{-5}$ and $\varepsilon = 10^{-10}$ are presented in Appendix B.

**Table 2**. Horizon of predictability for acceptable error $10^{-2}$ and different precisions $m = 500, 1000, \dots, 4500$. The estimation according to Eq. (11) is compared with the Analytic Computation Eq. (7) and the Recursive Computation Eq. (6).

| **Precision $m$** | **Horizon of Predictability Eq. (11)** | **Horizon of Predictability of the Analytic Computation $\tau_P^A$** | **Horizon of Predictability of the Recursive Computation $\tau_P^R$** |
|---|---|---|---|
| **500** | 1654 | 1659 | 1656 |
| **1000** | 3315 | 3322 | 3320 |
| **1500** | 4976 | 4980 | 4979 |
| **2000** | 6637 | 6642 | 6640 |
| **2500** | 8298 | 8302 | 8303 |
| **3000** | 9959 | 9965 | 9962 |
| **3500** | 11620 | 11624 | 11622 |
| **4000** | 13281 | 13286 | 13285 |
| **4500** | 14942 | 14947 | 14950 |

The theoretical value of the horizon of predictability, Eq. (11), is confirmed. The computed orbits remained within the acceptable error $\varepsilon = 10^{-2}$, Eq. (2). We observe that the horizon of predictability from Eq. (11) is shorter by 2 iterations on average than the corresponding horizons of predictability of the Analytic and Recursive Computations. This happens because the worst-case scenarios are considered (Proposition 1). Moreover, the observation that the Analytic Computation does not improve the Recursive Computation (Remark at the end of Section 5) is confirmed numerically.

The comparison of the orbits computed with the Error-Free Computation and with the Computations of the Logistic map with precision $m = 500$ are presented in Fig. 2. The exponential growth of the error is clearly depicted. The computations for more values of $m$ are presented in Appendix C.

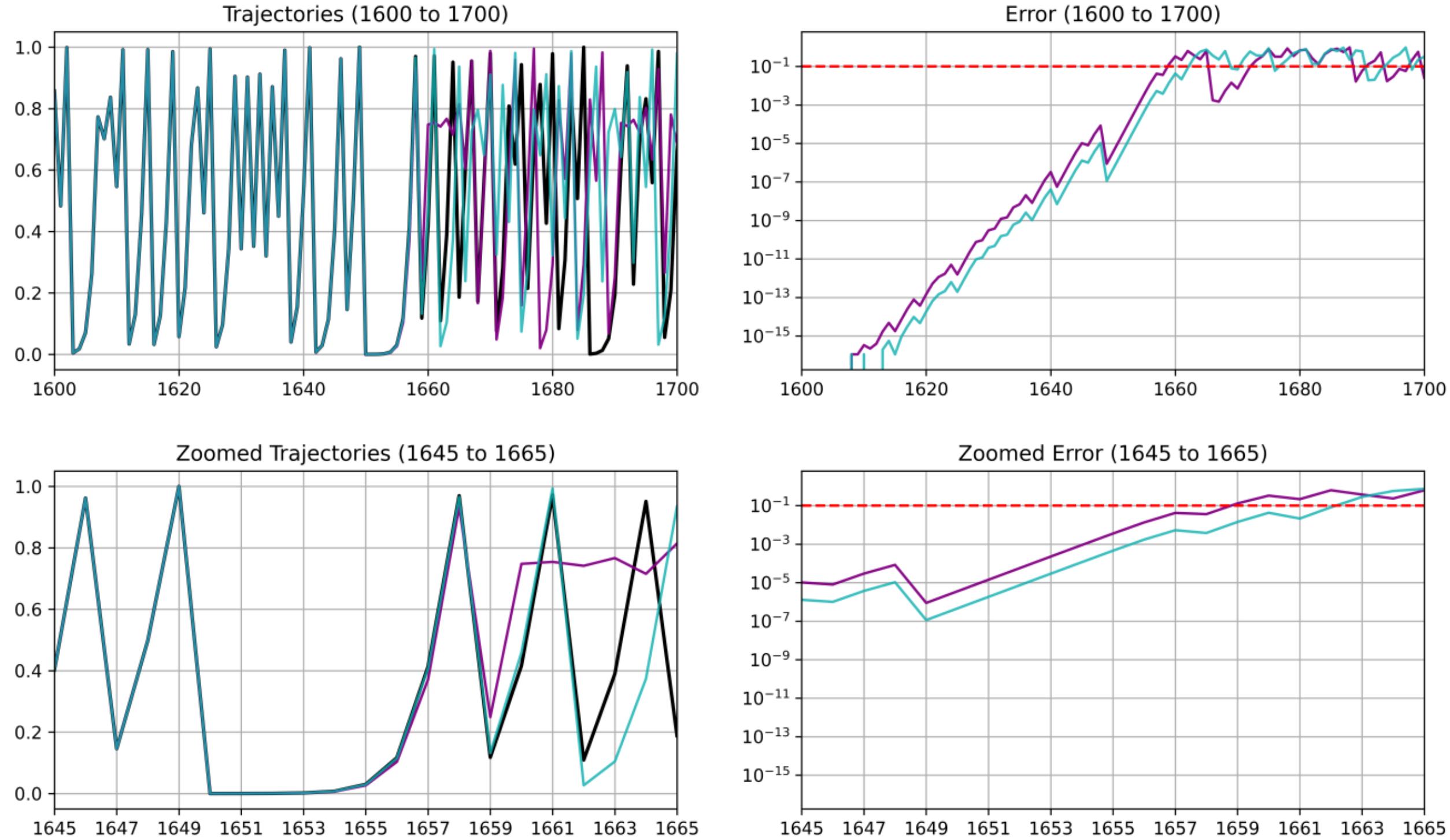

**Fig. 2.** Computations of the Logistic map with precision $m = 500$. The purple line is the Recursive Computation and the blue line is the Analytic Computation. The black line is the Error-Free Computation obtained via the transformation $g$, Eq. (9).

We observe that both the Recursive and Analytic Computations exhibit more or less indistinguishable behavior, confirming the remark at the end of Section 5.

## 7. Conclusions

Chaos revealed an inherent paradox in science. Mathematically deterministic models are in practice non-determinable. Every computational machine is inherently bound by finite precision. Beyond the horizon of predictability, computations are contaminated by errors and hallucinations are not even detectable. The computed sequences, beyond the horizon of predictability, are speculations devoid of interpretation and explanation. The estimation of the horizon of predictability is a way to avoid the paradox of predictability. In this perspective, we estimated the horizon of predictability and compared 3 main Computation strategies of the Logistic map, namely the Recursive Computation (Section 4), the Analytic Computation (Section 5), and the Error-Free Computation (Section 6). Specifically:

- We provide a way for generating error-free time series of the Logistic map (Section 6) which can serve as a "gold standard" for testing computation algorithms since they have effectively unlimited horizon of predictability.
- We introduced and validated a novel formula for the horizon of predictability of the Analytic Computation, Eq. (11), of Proposition 1, Section 5, which allows us to find how many iterations we can trust using specific precision and acceptable error. No estimation of the horizon of predictability for the Analytic Computation is found. Only estimations for the Recursive Computation [97], [98], [99] are available.
- It is surprising that the Analytic Computations are not more reliable than the Recursive Computations. This is explained in the Remark at the end of Section 5 and confirmed numerically in Section 6. However, constructing the Analytic Computations demands significantly more resources and time than the Recursive Computations (Appendix D).

**Significance of our findings**:

- The ability to compute a chaotic system allows to compare the performance of neural networks designed to predict chaotic time series which is a hot topic [101], [102], [103], [104], [105], [106], [107], [108].

- The Logistic map has been applied to encryption [109], [110]. However, due to finite numerical precision, the orbits of digital chaotic systems can degrade and become periodic, which compromises the security of chaos-based schemes [111], [112], [113]. Therefore, ensuring the reliability and numerical accuracy of chaotic orbit computations is crucial for maintaining the cryptographic strength of such systems.
- The extension of the horizon of predictability challenges the very limits of what can be "trusted" in applications involving chaotic economic models. In markets where price evolution is chaotic, the horizon of predictability (duration of reliable predictions) indicates the bound of rationality. Beyond the horizon of predictability, we have to resort to speculations. The extension of the horizon of predictability has profound implications for reliable forecasting of future prices in Economics [114], in Industrial Organization and Game Theory [115], as well as in Financial Markets [116], [117].

## Appendix A. Error-Free Computation of the Tent Map

Using the rational representation of the numbers in the unit interval [0,1] [34], the Tent map, Eq. (8), is:

$$y = \binom{\alpha}{\beta} \mapsto T(y) = \begin{cases} \binom{2\alpha}{\beta}, & \text{if } \frac{\alpha}{\beta} \leq \frac{1}{2} \\ \binom{2(\beta-\alpha)}{\beta}, & \text{if } \frac{\alpha}{\beta} > \frac{1}{2} \end{cases} \tag{A.1}$$

The numbers $y$ are a finite subset of rational numbers $\widetilde{\mathbb{Q}}$ [34]. In the rational representation each number $y$ is represented by a pair of relative prime integers. We denote by $\alpha$ the nominator and by $\beta$ the denominator. We observe from Eq. (A.1) that at each iteration of the Tent map the denominator $\beta$ does not change, while the nominator $\alpha$ changes to a number not greater than $\beta$.

In the rational representation, all numerical computations remain bounded withing the mantissa $m$, Eq. (1). The error-free computations are precisely the unstable periodic orbits of the Tent map, since the rational numbers define eventually periodic orbits. The period depends on the number $N$ of decimal digits of the initial point. Baba and Nagashima found [118] that any orbit starting from a point $y = 0.y_0 y_1 \dots y_{N-1}$ with $y_{N-1} \neq 0,5$ necessarily falls into one periodic orbit of period $2 \cdot 5^{N-1}$. This implies that an initial point of say 7 digits will generate a time series with a period of 31250, but before the iteration 31251, the sequence is indistinguishable from a chaotic orbit. In practice, we can construct orbits with arbitrarily large periods.

## Appendix B. Horizons of Predictability of the Logistic Map for Different Precisions

**Table B.1**. Horizon of predictability for acceptable error $10^{-5}$ and different precisions $m = 500, 1000, \dots, 4500$. The estimation according to Eq. (11) is compared with the Analytic Computation Eq. (7) and the Recursive Computation Eq. (6).

| **Precision** $m$ | **Horizon of Predictability Eq. (11)** | **Horizon of Predictability of the Analytic Computation** $\tau_P^A$ | **Horizon of Predictability of the Recursive Computation** $\tau_P^R$ |
|---|---|---|---|
| **500** | 1644 | 1648 | 1645 |
| **1000** | 3305 | 3312 | 3311 |
| **1500** | 4966 | 4970 | 4968 |
| **2000** | 6627 | 6631 | 6630 |
| **2500** | 8288 | 8292 | 8292 |
| **3000** | 9949 | 9955 | 9951 |
| **3500** | 11610 | 11614 | 11613 |
| **4000** | 13271 | 13277 | 13276 |
| **4500** | 14932 | 14937 | 14940 |

**Table B.2**. Horizon of predictability for acceptable error $10^{-10}$ and different precisions $m = 500, 1000, \dots, 4500$. The estimation according to Eq. (11) is compared with the Analytic Computation Eq. (7) and the Recursive Computation Eq. (6).

| Precision $m$ | Horizon of Predictability Eq. (11) | Horizon of Predictability of the Analytic Computation $\tau_P^A$ | Horizon of Predictability of the Recursive Computation $\tau_P^R$ |
|---|---|---|---|
| **500** | 1628 | 1632 | 1630 |
| **1000** | 3289 | 3300 | 3293 |
| **1500** | 4950 | 4954 | 4951 |
| **2000** | 6611 | 6615 | 6614 |
| **2500** | 8272 | 8275 | 8276 |
| **3000** | 9933 | 9939 | 9934 |
| **3500** | 11594 | 11598 | 11595 |
| **4000** | 13254 | 13260 | 13259 |
| **4500** | 14915 | 14920 | 14924 |

## Appendix C. Comparison of Computations of the Logistic Map

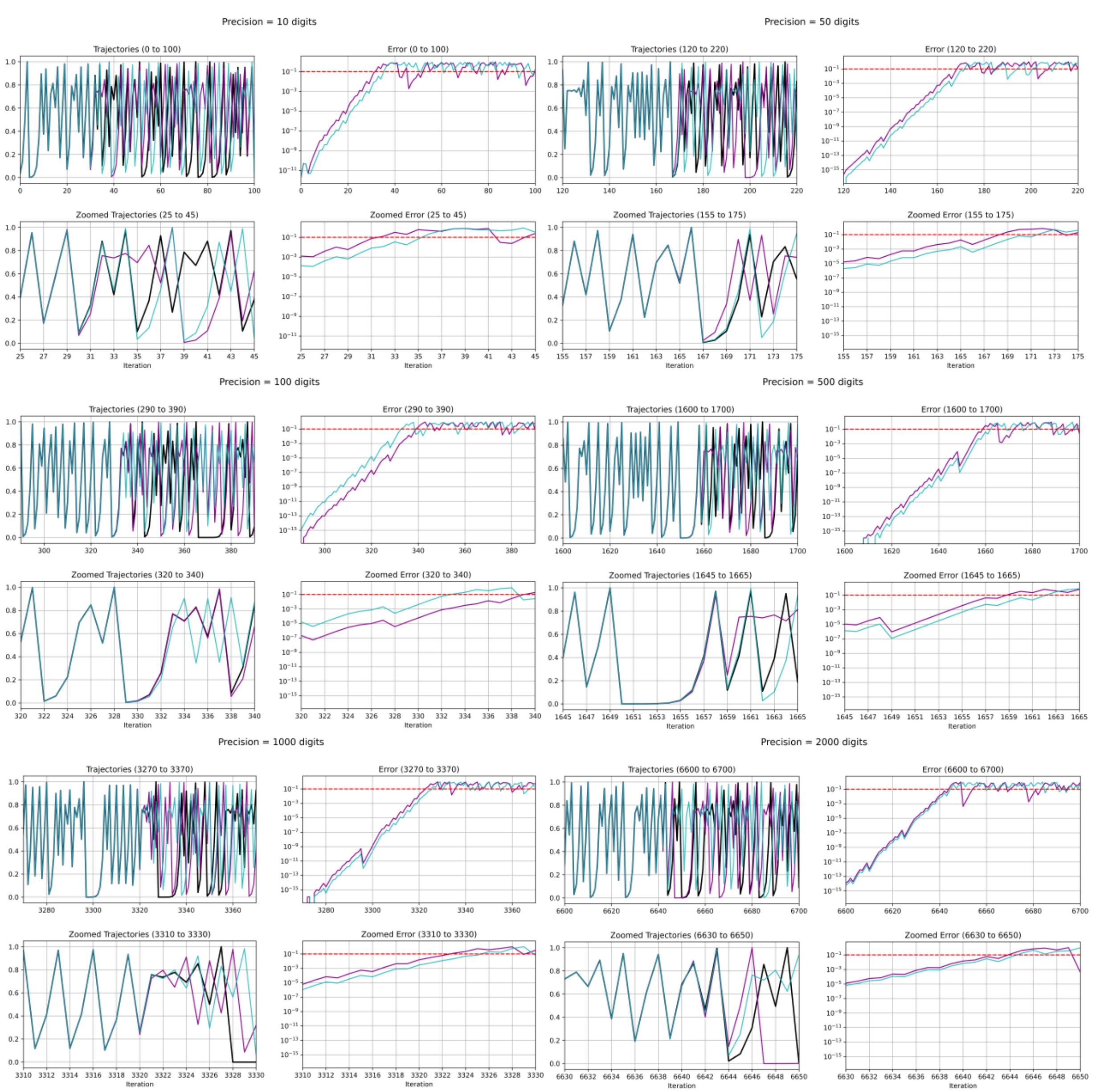

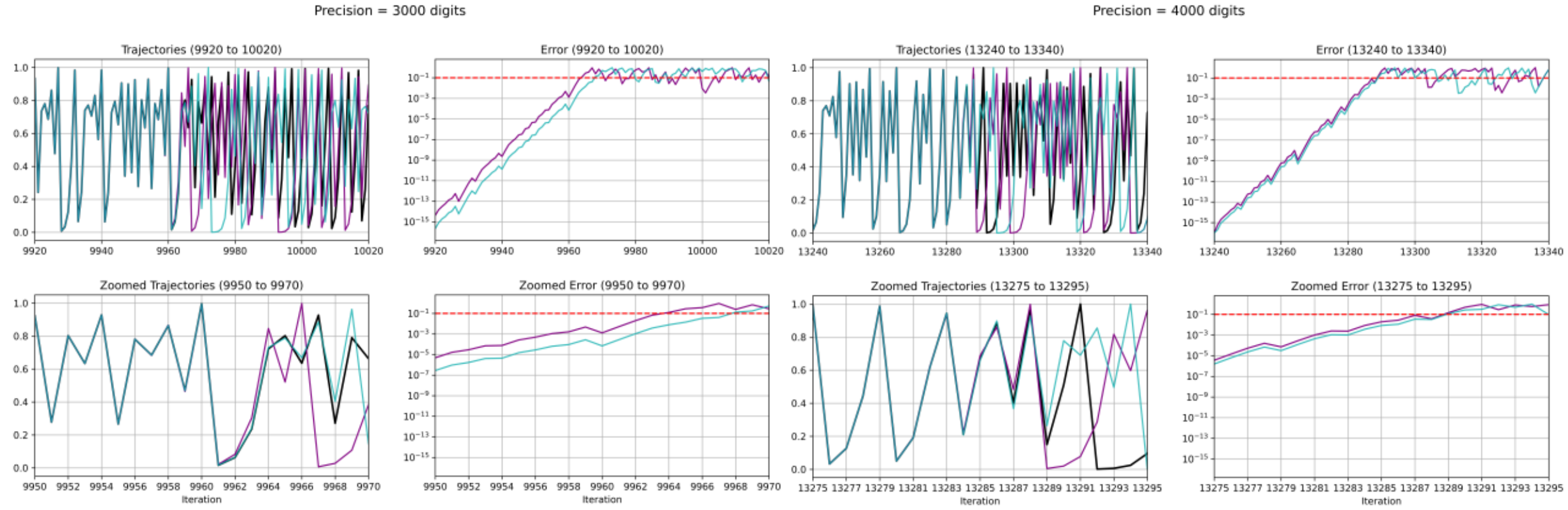

**Fig. C.1.** Computations of the Logistic map with precision $m = 10, 50, 100, 500, 1000, 2000, 3000, 4000$. The purple line is the Recursive Computation and the blue line is the Analytic Computation. The black line is the Error-Free Computation obtained via the transformation $g$, Eq. (9).

## Appendix D. Duration of Analytic and Recursive Computations

The time required to generate the time series using the Analytic Computation is significantly longer than the time of the Recursive Computation (Table D.1). In both cases the duration of computation grows exponentially with the precision $m$ (Fig. D.1).

**Table D.1**. The computation time of the Recursive and Analytic Computations for 10,000 iterations of the Logistic map with various precisions $m$.

| Precision $m$ | Analytic Computation Time (sec) | Recursive Computation Time (sec) |
|---|---|---|
| **10** | 0.5925 | 0.0560 |
| **50** | 0.6456 | 0.0591 |
| **100** | 0.7496 | 0.0661 |
| **500** | 1.9056 | 0.0976 |
| **1000** | 5.2105 | 0.1656 |
| **2000** | 17.1838 | 0.3663 |
| **3000** | 34.6498 | 0.6180 |
| **4000** | 64.6574 | 0.9803 |
| **5000** | 101.4753 | 1.4061 |
| **6000** | 132.6286 | 1.7080 |
| **7000** | 188.9581 | 2.2119 |
| **8000** | 245.5608 | 2.7399 |
| **9000** | 319.0776 | 3.3999 |
| **10000** | 401.6651 | 4.0490 |
| **11000** | 433.3804 | 4.3628 |
| **12000** | 518.4341 | 4.9142 |
| **13000** | 615.9455 | 5.8450 |
| **14000** | 741.0641 | 6.5411 |
| **15000** | 863.7403 | 7.2908 |

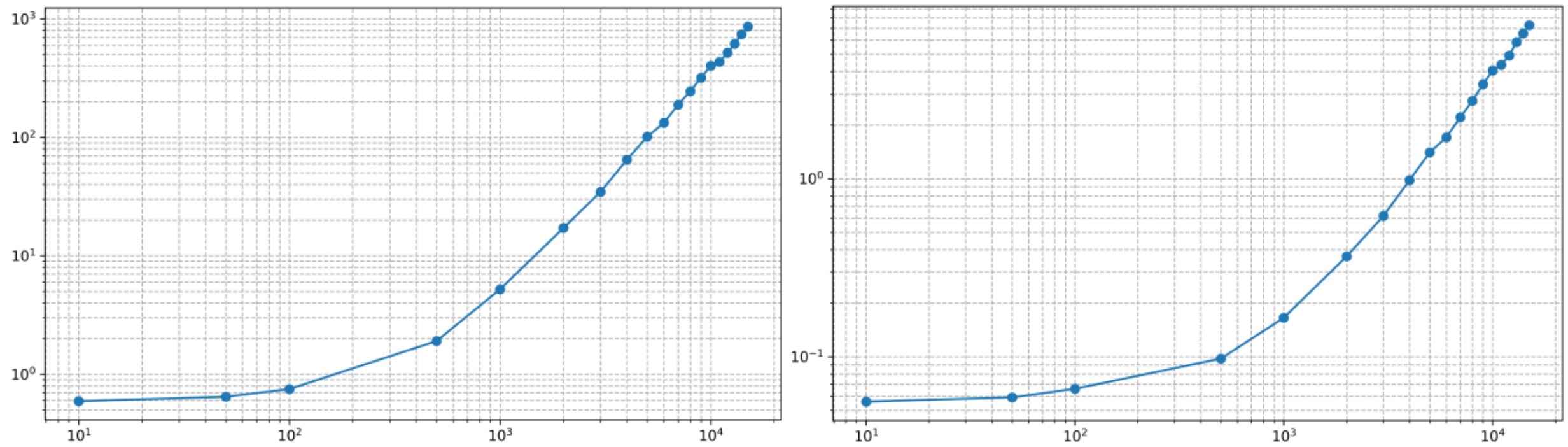

**Fig. D.1**. Duration of computation (vertical axis) versus precision $m$ (horizontal axis) in log-log scale of Analytic Computation (left) and Recursive Computation (right).

# References

[1] H. Poincaré, Les méthodes nouvelles de la mécanique céleste (Vol. 2), Gauthier-Villars et fils, imprimeurs-libraires, 1893.

[2] F. Lorenzelli, The Essence of Chaos, 1st ed.; CRC Press: London, UK, 1993.

[3] E. N. Lorenz, Computational chaos-a prelude to computational instability, Physica D: Nonlinear Phenomena, 35(3), 299-317. https://doi.org/10.1016/0167-2789(89)90072-9, 1989.

[4] E. N. Lorenz, Computational periodicity as observed in a simple system, Tellus A: Dynamic Meteorology and Oceanography, 58(5), 549-557. https://doi.org/10.1111/j.1600-0870.2006.00201.x, 2006.

[5] A. J. Lichtenberg and M. A. Lieberman, Regular and Chaotic Dynamics, Springer, New York. https://doi.org/10.1007/978-1-4757-2184-3, 1983.

[6] R. L. Devaney, An Introduction to Chaotic Dynamical Systems, 3rd ed.; CRC Press: Boca Raton, FL, USA, 2021.

[7] R. C. Hilborn, Chaos and Nonlinear Dynamics: An Introduction for Scientists and Engineers, 2nd ed.; Oxford University Press: New York, NY, USA, 2000.

[8] S. H. Strogatz, Nonlinear Dynamics and Chaos: With Applications to Physics, Biology, Chemistry, and Engineering, 2nd ed.; CRC Press: Boca Raton, FL, USA, 2015.

[9] J. P. Eckmann and D. Ruelle, Ergodic theory of chaos and strange attractors, Reviews of Modern Physics, 57(3), 617–656. https://doi.org/10.1103/RevModPhys.57.617, 1985.

[10] F. C. Moon, Chaotic and Fractal dynamics, A Wiley-Interscience Publication, John Wiley & Sons, Inc. New York, 1992.

[11] K. Alligood, T. Sauer and J. Yorke, Chaos, An Introduction to Dynamical Systems, Springer Science+Business Media. https://doi.org/10.1007/b97589, 1996.

[12] E. Ott, Chaos in Dynamical Systems, 2nd ed. Cambridge University Press, 2002.

[13] M. J. Lighthill, "The recently recognized failure of predictability in Newtonian dynamics," *Proceedings of the Royal Society of London. A. Mathematical and Physical Sciences,* Vols. 407. 1832: 35-50. https://doi.org/10.1098/rspa.1986.0082, 1986.

[14] S. Liao, On the reliability of computed chaotic solutions of non-linear differential equations, Tellus A: Dynamic Meteorology and Oceanography, 61(4), 550-564. https://doi.org/10.1111/j.1600-0870.2009.00402.x, 2008.

[15] M. Sangiorgio, F. Dercole and G. Guariso, Deep Learning in Multi-step Prediction of Chaotic Dynamics, From Deterministic Models to Real-World Systems, Springer, Cham, Switzerland. https://doi.org/10.1007/978-3-030-94482-7, 2021.

[16] A. Kolmogorov, New Metric Invariant of Transitive Dynamical Systems and Endomorphisms of Lebesgue Spaces, Doklady of Russian Academy of Sciences 119, N5, 861-864, 1958.

[17] P. Billingsley, Ergodic Theory and Information, J. Wiley New York, 1965.

[18] I. P. Cornfeld, S. F. Fomin and Y. G. Sinai, Ergodic Theory, Springer-Verlag, 1981.

[19] Y. Pesin, Characteristic Lyapunov Exponents and Smooth Ergodic Theory, Russian Math. Surveys 32, 4, 55-114, 1977.

[20] R. Mane, A proof of Pesin's formula, Ergodic Theory and Dynamical Systems 1, 95-102. http://doi.org/10.1017/S0143385700001188, 1981.

[21] V. Rokhlin, Exact endomorphisms of a Lebesgue space, Izv. Akad. Nauk SSSR Ser. Mat. 25, 499-530, 1961.

[22] W. Parry, On Rohlin's Formula For Entropy, Acta Mathematica Hungarica 15, 107-113, 1964.

[23] S. Li, When Chaos Meets Computers, arXiv: Chaotic Dynamics, 2004.

[24] H. Hu, Y. Xu and Z. Zhu, A method of improving the properties of digital chaotic system, Chaos, Solitons & Fractals, 38(2), 439-446. https://doi.org/10.1016/j.chaos.2006.11.027, 2008.

[25] S. Galatolo, M. Hoyrup and C. Rojas, Statistical properties of dynamical systems–simulation and abstract computation, Chaos, Solitons & Fractals, 45(1), 1-14. https://doi.org/10.1016/j.chaos.2011.09.011, 2012.

[26] S. Liao, On the numerical simulation of propagation of micro-level inherent uncertainty for chaotic dynamic systems, Chaos, Solitons & Fractals, 47, 1-12. https://doi.org/10.1016/j.chaos.2012.11.009, 2013.

[27] O. Mali, P. Neittaanmäki and S. Repin, Errors Arising in Computer Simulation Methods, In: Accuracy Verification Methods. Computational Methods in Applied Sciences, vol 32. Springer, Dordrecht. https://doi.org/10.1007/978-94-007-7581-7_1, 2014.

[28] L. De Micco, M. Antonelli and H. A. Larrondo, Stochastic degradation of the fixed-point version of 2D-chaotic maps, Chaos, Solitons & Fractals, 104, 477-484. https://doi.org/10.1016/j.chaos.2017.09.007, 2017.

[29] B. M. Boghosian, P. Coveney and H. Wang, A New Pathology in the Simulation of Chaotic Dynamical Systems on Digital Computers, Adv Theory Simul. 2(12):1900125. https://doi.org/10.1002/adts.201900125, 2019.

[30] R. A. Elmanfaloty and E. Abou-Bakr, Random property enhancement of a 1D chaotic PRNG with finite precision implementation, Chaos, Solitons & Fractals, 118, 134-144. https://doi.org/10.1016/j.chaos.2018.11.019, 2019.

[31] J. Zheng, H. Hu, H. Ming and X. Liu, Theoretical design and circuit implementation of novel digital chaotic systems via hybrid control, Chaos, Solitons & Fractals, 138, 109863. https://doi.org/10.1016/j.chaos.2020.109863, 2020.

[32] C. Fan and Q. Ding, Design and geometric control of polynomial chaotic maps with any desired positive Lyapunov exponents, Chaos, Solitons & Fractals, 169, 113258. https://doi.org/10.1016/j.chaos.2023.113258, 2023.

[33] P. V. Coveney, Sharkovskii's theorem and the limits of digital computers for the simulation of chaotic dynamical systems, Journal of Computational Science, 83, 102449. https://doi.org/10.1016/j.jocs.2024.102449, 2024.

[34] R. T. Gregory and E. V. Krishnamurthy, Methods and Applications of Error-Free Computation, Springer-Verlag, New York, Berlin, Heidelberg, Tokyo, 1984.

[35] P. M. Binder and R. V. Jensen, Simulating chaotic behavior with finite-state machines, Physical Review A, 34(5), 4460. https://doi.org/10.1103/PhysRevA.34.4460, 1986.

[36] C. Beck and G. Roepstorff, Effects of phase space discretization on the long-time behavior of dynamical systems, Physica D: Nonlinear Phenomena, 25(1-3), 173-180. https://doi.org/10.1016/0167-2789(87)90100-X, 1987.

[37] C. Grebogi, E. Ott and J. A. Yorke, Roundoff-induced periodicity and the correlation dimension of chaotic attractors, Physical Review A, 38(7), 3688. https://doi.org/10.1103/PhysRevA.38.3688, 1988.

[38] M. Blank, Pathologies generated by round-off in dynamical systems, Physica D: Nonlinear Phenomena, 78(1-2), 93-114. https://doi.org/10.1016/0167-2789(94)00103-0, 1994.

[39] P. Diamond, P. Kloeden and A. Pokrovskii, An invariant measure arising in computer simulation of a chaotic dynamical system, J Nonlinear Sci 4, 59–68. https://doi.org/10.1007/BF02430627, 1994.

[40] P. Diamond, P. Kloeden, A. Pokrovskii and A. Vladimirov, Collapsing effects in numerical simulation of a class of chaotic dynamical systems and random mappings with a single attracting centre, Physica D: Nonlinear Phenomena, 86(4), 559-571. https://doi.org/10.1016/0167-2789(95)00188-A, 1995.

[41] A. Iglesias, J. M. Gutiérrez, J. Güémez and M. A. Matías, Chaos suppression through changes in the system variables and numerical rounding errors, Chaos, Solitons & Fractals, 7(8), 1305-1316. https://doi.org/10.1016/0960-0779(95)00072-0, 1996.

[42] S. Li, G. Chen and X. Mou, On the dynamical degradation of digital piecewise linear chaotic maps, International journal of Bifurcation and Chaos, 15(10), 3119-3151. https://doi.org/10.1142/S0218127405014052, 2005.

[43] T. Hu and S. Liao, On the risks of using double precision in numerical simulations of spatio-temporal chaos, Journal of Computational Physics, 418, 109629. https://doi.org/10.1016/j.jcp.2020.109629, 2020.

[44] A. V. Tutueva, E. G. Nepomuceno, A. I. Karimov, V. S. Andreev and D. N. Butusov, Adaptive chaotic maps and their application to pseudo-random numbers generation, Chaos, Solitons & Fractals, 133, 109615. https://doi.org/10.1016/j.chaos.2020.109615, 2020.

[45] B. M. El-Den, S. Aldosary, H. Khaled, T. M. Hassan and W. Raslan, Leveraging Finite-Precision Errors in Chaotic Systems for Enhanced Image Encryption, In IEEE Access, vol. 12, pp. 176057-176069, https://doi.org/10.1109/ACCESS.2024.3462807, 2024.

[46] I. Prigogine, From Being to Becoming, Freeman, New York, 1980.

[47] I. Antoniou and S. Tasaki, Generalized Spectral Decompositions of Mixing Dynamical Systems, International J. of Quantum Chemistry 46, 425-474, 1993.

[48] I. Antoniou and S. Tasaki, Spectral Decompositions of the Renyi Map, J. Phys. A: Math. Gen 26, 73-94, 1993.

[49] S. Tasaki, I. Antoniou and Z. Suchanecki, Spectral Decomposition and Fractal Eigenvectors for a Class of Piecewise Linear Maps, Chaos, Solitons and Fractals 4, 227-254. https://doi.org/10.1016/0960-0779(94)90147-3, 1994.

[50] I. Antoniou and B. Qiao, Spectral Decomposition of the Tent Maps and the Isomorphism of Dynamical Systems, Phys. Lett. A215, 280-290. https://doi.org/10.1016/0375-9601(96)00104-1, 1996.

[51] B. Qiao and I. Antoniou, Spectral Decomposition of the Chebyshev Maps, Physica A233, 449-457. https://doi.org/10.1016/S0378-4371(96)00219-1, 1996.

[52] Z. Suchanecki, I. Antoniou, S. Tasaki and O. Bantlow, Rigged Hilbert Spaces for Chaotic Dynamical Systems, J. Math. Phys. 37 (11): 5837-5847. https://doi.org/10.1063/1.531703, 1996.

[53] I. Antoniou and B. Qiao, Spectral decomposition of the chaotic logistic map, Nonlinear World 4, 135-143, 1997.

[54] I. Antoniou, B. Qiao and Z. Suchanecki, Generalized Spectral Decomposition and Intrinsic Irreversibility of the Arnold Cat Map, Chaos Solitons and Fractals 8, 77 – 90. https://doi.org/10.1016/S0960-0779(96)00056-2, 1997.

[55] O. F. Brandtlow, I. Antoniou and Z. Suchanecki, Resonances of Dynamical Systems and Fredholm-Riesz Operators on Rigged Hilbert Space, Computers Math. Applic. 34, 95-102. https://doi.org/10.1016/S0898-1221(97)00119-3, 1997.

[56] I. Antoniou and Z. Suchanecki, The Fuzzy Logic of Chaos and Probabilistic Inference, Foundations of Physics 27, 333 – 362. https://doi.org/10.1007/BF02550161, 1997.

[57] I. Antoniou and K. Gustafson, From Irreversible Markov Semigroups to Chaotic Dynamics, Physica A236, 296-308. https://doi.org/10.1016/S0378-4371(96)00375-5, 1997.

[58] I. Antoniou, K. Gustafson and Z. Suchanecki, On the Inverse Problem of Statistical Physics: From Irreversible Semigroups to Chaotic Dynamics, Physica A 252, 345-361. https://doi.org/10.1016/S0378-4371(97)00622-5, 1998.

[59] I. Antoniou, V. Sadovnichii and S. Shkarin, Time Operators and Shift Representation of Dynamical Systems, Physica A299, 299-313. https://doi.org/10.1016/S0378-4371(99)00070-9, 1999.

[60] I. Antoniou, Y. Melnikov, S. Shkarin and Z. Suchanecki, Extended Spectral Decompositions of the Renyi Map, Chaos, Solitons and Fractals 11, 393-421. https://doi.org/10.1016/S0960-0779(98)00309-9, 2000 .

[61] E. Kalnay, Atmospheric Modeling, Data Assimilation and Predictability, Cambridge: Cambridge University Press. https://doi.org/10.1017/CBO9780511802270, 2002.

[62] G. Evensen, The Ensemble Kalman Filter: theoretical formulation and practical implementation, Ocean Dynamics 53, 343–367. https://doi.org/10.1007/s10236-003-0036-9, 2003.

[63] A. Lasota and M. C. Mackey, Chaos, fractals, and noise: stochastic aspects of dynamics, (Vol. 97). Springer Science & Business Media., 2013.

[64] I. Antoniou, I. Gialampoukidis and E. Ioannidis, Age and Time Operator of Evolutionary Processes, In: Atmanspacher, H., Filk, T., Pothos, E. (eds) Quantum Interaction. QI 2015. Lecture Notes in Computer Science(), vol 9535. Springer, Cham. https://doi.org/10.1007/978-3-319-28675-4_5, 2016.

[65] A. K. Angelidis, K. Goulas, C. Bratsas, G. C. Makris, M. P. Hanias, S. G. Stavrinides and I. E. Antoniou, Distinction of Chaos from Randomness Is Not Possible from the Degree Distribution of the Visibility and Phase Space Reconstruction Graph, Entropy, 26(4), 341. https://doi.org/10.3390/e26040341, 2024.

[66] H. A. Simon, Models of man: social and rational; mathematical essays on rational human behavior in society setting, New York: Wiley., 1957.

[67] M. Klaes and E. M. Sent, A Conceptual History of the Emergence of Bounded Rationality, History of Political Economy, 37(1):27–59. https://doi.org/10.1215/00182702-37-1-27, 2005.

[68] N. C. A. da Costa and F. A. Doria, Undecidability and incompleteness in classical mechanics, Int J Theor Phys 30, 1041–1073. https://doi.org/10.1007/BF00671484, 1991.

[69] M. B. Pour-El and J. I. Richards, Computability in analysis and physics, (Vol. 1). Cambridge University Press. https://doi.org/10.1017/9781316717325, 2017.

[70] S. M. Ulam and J. von Neumann, On Combination of Stochastic and Deterministic Processes, Bull. Amer. Math. Soc. 53, 1120, 1947.

[71] P. R. Stein and S. M. Ulam, Non-linear transformation studies on electronic computers, Rozprawy Matematyczene 39, 1-66., 1964.

[72] T. Tsuchiya, A. Szabo and N. Saitô, "Exact Solutions of Simple Nonlinear Difference Equation Systems that show Chaotic Behavior," *Zeitschrift für Naturforschung A,* Vols. 38. 9. 1035-1039. https://doi.org/10.1515/zna-1983-0913, 1983.

[73] S. F. W. Katsura, "Exactly solvable models showing chaotic behavior," *Physica A: Statistical Mechanics and its Applications,* Vols. 130(3), 597-605. https://doi.org/10.1016/0378-4371(85)90048-2, 1985.

[74] M. A. García-Ñustes, E. Hernández-García and J. A. González, Universal functions and exactly solvable chaotic systems, São Paulo Journal of Mathematical Sciences, 2(2), 204-221. https://doi.org/10.11606/issn.2316-9028.v2i2p204-221, 2008.

[75] R. M. Corless, C. Essex and M. A. H. Nerenberg, Numerical methods can suppress chaos, Physics Letters A, 157(1), 27-36. https://doi.org/10.1016/0375-9601(91)90404-V, 1991.

[76] R. M. Corless, What good are numerical simulations of chaotic dynamical systems?, Computers & Mathematics with Applications, 28(10-12), 107-121. https://doi.org/10.1016/0898-1221(94)00188-X, 1994.

[77] L.-S. Yao, Computed chaos or numerical errors, Nonlinear Analysis: Modelling and Control, 15(1), pp. 109–126. https://doi.org/10.15388/NA.2010.15.1.14368, 2010.

[78] V. E. Isaacson and H. Keller, Analysis of Numerical Methods, Dover, New Yorl edition of the original 1966 edition, Wiley, New York, 1993.

[79] R. Lozi, Can we trust in numerical computations of chaotic solutions of dynamical systems?, In Topology and dynamics of Chaos: In celebration of Robert Gilmore's 70th birthday (pp. 63-98). https://doi.org/10.1142/9789814434867_0004, 2013.

[80] S. Qin and S. Liao, Influence of numerical noises on computer-generated simulation of spatio-temporal chaos, Chaos, Solitons & Fractals, 136, 109790. https://doi.org/10.1016/j.chaos.2020.109790, 2020.

[81] A. Sarkovskii, Coexistence of cycles of a continuous map of a line to itself, Ukr. Mat. Z., 16, 61-71, 1964.

[82] K. Burns and B. Hasselblatt, The Sharkovsky Theorem: A Natural Direct Proof, The American Mathematical Monthly, 118(3), 229–244. https://doi.org/10.4169/amer.math.monthly.118.03.229, 2011.

[83] R. May, Simple mathematical models with very complicated dynamics, Nature 261, 459–467. https://doi.org/10.1038/261459a0, 1976.

[84] E. A. Jackson, Perspectives of Nonlinear Dynamics, vol. 1, Cambridge University Press, Cambridge, 1989.

[85] S. Rabinovich, G. Berkolaiko and S. Havlin, Solving nonlinear recursions, Journal of Mathematical Physics, 37(11): 5828-5836. https://doi.org/10.1063/1.531702, 1996.

[86] M. W. Hirsch, S. Smale and R. L. Devaney, Differential Equations, Dynamical Systems, and an Introduction to Chaos, 3rd ed.; Academic Press: Cambridge, MA, USA, 2013.

[87] A. T. Ruslan, Marwan and Q. Aini, Behavior of logistic map and some of its conjugate maps, AIP Conf. Proc. 2641 (1): 020002. https://doi.org/10.1063/5.0115103, 2022.

[88] G. Layek, Conjugacy of Maps, In: An Introduction to Dynamical Systems and Chaos. University Texts in the Mathematical Sciences. Springer, Singapore. https://doi.org/10.1007/978-981-99-7695-9_11, 2024.

[89] A. Saito and S. Ito, Computation of true chaotic orbits using cubic irrationals, Physica D: Nonlinear Phenomena, 268, 100-105. https://doi.org/10.1016/j.physd.2013.11.003, 2014.

[90] K. J. Persohn and R. J. Povinelli, Analyzing logistic map pseudorandom number generators for periodicity induced by finite precision floating-point representation, Chaos, Solitons & Fractals, 45(3), 238-245. https://doi.org/10.1016/j.chaos.2011.12.006, 2012.

[91] Z. Galias, Periodic orbits of the logistic map in single and double precision implementations, IEEE Transactions on Circuits and Systems II: Express Briefs, 68(11), 3471-3475. https://doi.org/10.1109/TCSII.2021.3081604, 2021.

[92] M. Klöwer, P. V. Coveney, E. A. Paxton and T. N. Palmer, Periodic orbits in chaotic systems simulated at low precision, Sci Rep 13, 11410. https://doi.org/10.1038/s41598-023-37004-4, 2023.

[93] J. Liu, H. Zhang and D. Song, The Property of Chaotic Orbits with Lower Positions of Numerical Solutions in the Logistic Map, Entropy. 16(11):5618-5632. https://doi.org/10.3390/e16115618, 2014.

[94] J. Valle and O. M. Bruno, Dynamics and patterns of the least significant digits of the infinite-arithmetic precision logistic map orbits, Chaos, Solitons & Fractals, 180, 114488. https://doi.org/10.1016/j.chaos.2024.114488, 2024.

[95] J. A. Oteo and J. Ros, Double precision errors in the logistic map: Statistical study and dynamical interpretation, Physical Review E—Statistical, Nonlinear, and Soft Matter Physics, 76(3), 036214. https://doi.org/10.1103/PhysRevE.76.036214, 2007.

[96] J. Machicao and O. M. Bruno, Improving the pseudo-randomness properties of chaotic maps using deep-zoom., Chaos: an interdisciplinary journal of nonlinear science, 27(5). https://doi.org/10.1063/1.4983836, 2017.

[97] P. Wang and X. Pan, The reliable solution and computation time of variable parameters logistic model, Theor Appl Climatol 132, 851–855. https://doi.org/10.1007/s00704-017-2136-3, 2018.

[98] E. G. Nepomuceno and E. M. Mendes, On the analysis of pseudo-orbits of continuous chaotic nonlinear systems simulated using discretization schemes in a digital computer, Chaos, Solitons & Fractals, 95, 21-32. https://doi.org/10.1016/j.chaos.2016.12.002, 2017.

[99] M. L. Peixoto, E. G. Nepomuceno, S. A. Martins and M. J. Lacerda, Computation of the largest positive Lyapunov exponent using rounding mode and recursive least square algorithm, Chaos, Solitons & Fractals, 112, 36-43. https://doi.org/10.1016/j.chaos.2018.04.032, 2018.

[100] P. Akritas, I. Antoniou and V. V. Ivanov, Identification and prediction of discrete chaotic maps applying a Chebyshev neural network, Chaos, Solitons & Fractals, 11(1-3), 337-344. https://doi.org/10.1016/S0960-0779(98)00302-6, 2000.

[101] M. Sangiorgio and F. Dercole, Robustness of LSTM neural networks for multi-step forecasting of chaotic time series., Chaos, Solitons & Fractals, 139, 110045. https://doi.org/10.1016/j.chaos.2020.110045, 2020.

[102] L. Tang, Y. Bai, J. Yang and Y. Lu, A hybrid prediction method based on empirical mode decomposition and multiple model fusion for chaotic time series, Chaos, Solitons & Fractals, 141, 110366. https://doi.org/10.1016/j.chaos.2020.110366, 2020.

[103] M. Sangiorgio, F. Dercole and G. Guariso, Forecasting of noisy chaotic systems with deep neural networks, Chaos, Solitons & Fractals, 153, 111570. https://doi.org/10.1016/j.chaos.2021.111570, 2021.

[104] W. Cheng, Y. Wang, Z. Peng, X. Ren, Y. Shuai, S. Zang, H. Liu, H. Cheng and J. Wu, High-efficiency chaotic time series prediction based on time convolution neural network, Chaos, Solitons & Fractals, 152, 111304. https://doi.org/10.1016/j.chaos.2021.111304, 2021.

[105] G. Uribarri and G. B. Mindlin, Dynamical time series embeddings in recurrent neural networks, Chaos, Solitons & Fractals, 154, 111612. https://doi.org/10.1016/j.chaos.2021.111612, 2022.

[106] Y. Sun, L. Zhang and M. Yao, Chaotic time series prediction of nonlinear systems based on various neural network models, Chaos, Solitons & Fractals, 175, 113971. https://doi.org/10.1016/j.chaos.2023.113971, 2023.

[107] Q. Wang, L. Jiang, L. Yan, X. He, J. Feng, W. Pan and B. Luo, Chaotic time series prediction based on physics-informed neural operator, Chaos, Solitons & Fractals, 186, 115326. https://doi.org/10.1016/j.chaos.2024.115326, 2024.

[108] J. Valle and O. M. Bruno, Forecasting chaotic time series: Comparative performance of LSTM-based and Transformer-based neural network, Chaos, Solitons & Fractals, 192, 116034. https://doi.org/10.1016/j.chaos.2025.116034, 2025.

[109] A. Kanso and N. Smaoui, Logistic chaotic maps for binary numbers generations, Chaos, Solitons & Fractals, 40(5), 2557-2568. https://doi.org/10.1016/j.chaos.2007.10.049, 2009.

[110] L. Moysis, A. Tutueva, C. Volos, D. Butusov, J. M. Munoz-Pacheco and H. Nistazakis, A Two-Parameter Modified Logistic Map and Its Application to Random Bit Generation, Symmetry, 12, 829. https://doi.org/10.3390/sym12050829, 2020.

[111] D. D. Wheeler, Problems with chaotic cryptosystems, Cryptologia, 13(3), 243-250. https://doi.org/10.1080/0161-118991863934, 1989.

[112] S. Li, X. Mou, Y. Cai, Z. Ji and J. Zhang, On the security of a chaotic encryption scheme: problems with computerized chaos in finite computing precision, Computer physics communications, 153(1), 52-58. https://doi.org/10.1016/S0010-4655(02)00875-5, 2003.

[113] F. Sun, S. Liu, Z. Li and Z. Lü, A novel image encryption scheme based on spatial chaos map, Chaos, Solitons & Fractals, 38(3), 631-640. https://doi.org/10.1016/j.chaos.2008.01.028, 2008.

[114] R. V. Jensen and R. Urban, Chaotic price behavior in a non-linear cobweb model, Economics Letters, 15(3-4), 235-240. https://doi.org/10.1016/0165-1765(84)90106-X, 1984.

[115] M. Kopel, Simple and complex adjustment dynamics in Cournot duopoly models, Chaos, Solitons & Fractals, 7(12), 2031-2048. https://doi.org/10.1016/S0960-0779(96)00070-7, 1996.

[116] R. H. Day and W. Huang, Bulls, bears and market sheep, Journal of Economic Behavior & Organization, 14(3), 299-329. https://doi.org/10.1016/0167-2681(90)90061-H, 1990.

[117] S. Sordi, A. Naimzada and M. J. Davila-Fernandez, A dynamic model of real-financial markets interaction, Economic Modelling, 107103. https://doi.org/10.1016/j.econmod.2025.107103, 2025.

[118] Y. Baba and H. Nagashima, A note on a class of periodic orbits of the tent-map, Progress of theoretical physics, 81(3), 541-543. https://doi.org/10.1143/PTP.81.541, 1989.